\documentclass[12p]{article}

\usepackage{amsmath}
\usepackage{amsfonts}
\usepackage{amssymb}
\usepackage{epsfig}
\usepackage{graphics}
\usepackage{subfig}

\usepackage[a4paper, margin=3cm]{geometry}

\newtheorem{proposition}{Proposition}[section]

\begin{document}

\title{Exactly Solvable Nonhomogeneous Burgers Equations with Variable Coefficients}

\maketitle

\begin{center}
{\bf \c{S}irin A. B\"{u}y\"{u}ka\c{s}{\i}k, Oktay K. Pashaev}\\
Dept. of Mathematics, Izmir Institute of Technology, \\
35430 Urla, Izmir, Turkey\\
sirinatilgan@iyte.edu.tr, oktaypashaev@iyte.edu.tr
\end{center}

\begin{abstract}
 We consider a nonhomogeneous Burgers equation with time variable coefficients of the form $\,\,U_{t}+(\dot{\mu}(t)/\mu(t))U+UU_{x}=(1/2\mu(t))U_{xx}-\omega^{2}(t)x,$ and  obtain an explicit  solution of the  general initial value problem in terms of  solution to a corresponding linear ODE. Special exact solutions such as generalized shock and multi-shock solitary waves, triangular wave, \emph{N}-wave  and rational type solutions are found and discussed. As exactly solvable models, we study forced Burgers equations with constant damping and an exponentially decaying diffusion coefficient. Different type of exact solutions are obtained for the critical, over and under damping cases, and their behavior is illustrated explicitly. In particular, the existence of inelastic type of collisions is observed by constructing multi-shock solitary wave solutions, and for the rational type  solutions the motion of the pole singularities is described.
\end{abstract}

\section{Introduction}

The nonlinear diffusion equation, known as Burgers equation (BE) after the extensive work of J. M. Burgers, \cite{Burgers1,Burgers} is  an important model which appears in various fields of physical science. In hydrodynamics, it is a standard  model of turbulence used to study propagation of nonlinear waves and shock formation, \cite{Bec}. It is used also to describe processes in gas dynamics \cite{Cole,Hopf},  nonlinear acoustics \cite{Lighthill}, heat conduction, and plasma physics.  In cosmology, the Burgers equation  is a good approximation to understand the formation  and distribution of matter at large scales, \cite{Woyczynski}.

The standard  Burgers equation  is of the form $ V_{t}+ V V_{x}=\nu V_{xx},$ where $V$ mostly represents the velocity field, $t$ is a time variable, $x\in \mathbb{R}$ is  the space variable and $\nu$ is a constant viscosity or diffusion coefficient. This equation is probably the simplest nonlinear model admitting direct linearization, and thus being C-integrable in contrast to S-integrable systems which require spectral transform technics. Indeed, exact explicit solutions of the Burgers equation can be obtained  by the Cole-Hopf transformation,  which transforms the nonlinear Burgers equation  to a linear heat equation,  \cite{Cole,Hopf}. Beside that, simple Ansatz method was also applied to find special solutions, like traveling waves and similarity solutions. Lately, other methods like, Hirota's direct method and  B\"{a}cklund transformation \cite{Wang}, hyperbolic function method \cite{Guixu}, and homogeneous balance method \cite{Bai-Zhao} were used  to construct new exact solutions of the BE.

As known, the main features of the Burgers equation are due to the simultaneous existence  of a nonlinear term and a linear diffusion term. If the diffusion is dominant over nonlinearity, the solution of the BE  approaches the solution of the diffusion  equation. On the other hand, if the nonlinear term dominates over the diffusion, one may expect formation of shock discontinuities. An interesting property of the BE appears when a balance occurs between the nonlinear effect and the effects of dissipative  nature. In that case, the system exhibits shock profile solitary wave solutions. Moreover, it is known that  Burgers equation has also multi-shock solitary wave solutions \cite{Whitham,Wang}. In that case, shocks of different amplitude and speed can fuse (merge) to a single shock, so that completely non-elastic interactions  may occur. Another important property of the Burgers equation is related with the rational type solutions. It is well known that, the zeros of the heat equation solution lead to pole singularities for the Burgers solution. Choodnovsky brothers \cite{Choodnovsky} and F. Calogero \cite{Calogero},  showed that the motion of the poles corresponds formally to the motion of one-dimensional particles interacting via simple two-body potentials, such that the corresponding many body problems are integrable. For recent work on the pole dynamics of the standard Burgers equation one can see \cite{Deconinck}.

 The standard BE, as mentioned above, is a well known exactly solvable model. However,  the nonhomogeneous and variable parametric Burgers equation, in general, is not integrable and very few exactly solvable models are known. For example, in case $ U_{t}+ U U_{x}=\nu U_{xx}+F(x,t),$
if the forcing term depends only on time, i.e. $F(x,t)=G(t)$, this equation can be transformed to a standard Burgers equation, see \cite{Orlowsky}. The IVP  with an elastic forcing term $F(x,t)=-k^2x+f(t)$ is discussed and analytic solutions are obtained in \cite{Vallee}. Later, the problem with $F(x,t)=G(t)x ,$ where $G(t)$ is arbitrary, was completely solved in terms of solution to the standard Burgers equation, see \cite{Eule}.  In \cite{Axel}, an invertible transformation between the nonhomogeneous BE and the stationary Schr\"{o}dinger equation was constructed so that each solution of the stationary Schr\"{o}dinger equation generated a fully time-dependent solution of the nonhomogeneous BE. Recently,  exact solutions  were obtained using the Cole-Hopf transformation and the Green's function approach, see \cite{Zola}. Transformation properties of a variable-coefficient Burgers equation were discussed in \cite{Sophocleous}. In \cite{Xu}, a forced Burgers model with space- and time-dependent coefficients of the form  $ U_{t}+ a(x,t)U U_{x}=b(x,t) U_{xx}+F(x,t),$ was investigated using a generalized Cole-Hopf transform and symbolic computation. For the significance of the generalized forced Burgers models and possible applications in various fields one can see again the discussion in \cite{Xu} and references given there.

    In this work, we consider  a nonhomogeneous Burgers equation (NHBE) with time variable coefficients of the form
\begin{equation}\label{burgers}
 U_{t}+\frac{\dot{\mu}(t)}{\mu(t)}U+U U_{x}=\frac{1}{2\mu(t)}U_{xx}-\omega^{2}(t)x\,,\,\,\,\,\,\,\,\,\,\,-\infty<x<\infty
\end{equation}
where $\Gamma(t)=\dot{\mu}(t)/\mu(t)$  is the damping term,  $D(t)=1/2\mu(t)$ is the diffusion coefficient, and $F(x,t)=-\omega^{2}(t)x$ is the forcing term which is linear in the space  variable $x.$ In Sec.2, we show that  solutions of the NHBE (\ref{burgers}) can be obtained in terms of solutions to the standard BE or heat equation and a related linear ODE.  As a result,  an explicit solution for the IVP of the NHBE (\ref{burgers}) is found in terms of  solution to the corresponding second order linear ODE
with variable frequency and damping. Then, some particular exact solutions such as  shock and multi-shock solitary type waves, triangular wave, \emph{N}-wave and rational type solutions are obtained.  In Sec.3, for comparative reasons, first we recall some solutions of the nonhomogeneous Burgers equation with constant coefficients. Then,
 exactly solvable  NHBE models (\ref{burgers}) with positive constant damping and exponentially decaying diffusion coefficient are considered. Different type of exact solutions mentioned in Sec.2 are obtained for the critical, under and over damping cases. We observe generalized traveling wave solutions which speed, steepness, and shock amplitude are functions of time. Special properties like interaction of shocks in multi-shock solitary solutions, and motion of pole singularities of rational type solutions are described explicitly.  Sec.4 includes brief summary and future plans.

\section{Nonhomogeneous Burgers equation with time dependent coefficients}

In the following proposition we obtain relation between  solutions of the nonhomogeneous variable coefficients Burgers equation and the standard Burgers equation. Then, Cole-Hopf transform allows us to find an explicit solution of the IVP for the variable coefficient NHBE (\ref{burgers}) in terms of solution to a corresponding second order linear ODE.

\begin{proposition} If $\,r(t)$ is  solution of the IVP for the linear ODE
\begin{equation}\label{clas-eq1}
\ddot{r}+\frac{\dot{\mu}(t)}{\mu(t)}\dot{r}+\omega^{2}(t)r=0,\,\,\,r(t_{0})=r_{0}\neq
0,\,\,\,\dot{r}(t_{0})=0,
\end{equation}
 then the IVP for the NHBE with variable coefficients
\begin{eqnarray}\label{burgers-ic1}
\left\{
\begin{array}{ll}
  U_{t}+\frac{\dot{\mu}(t)}{\mu(t)}U+U U_{x}=\frac{1}{2\mu(t)}U_{xx}-\omega^{2}(t)x\,,\\
U(x,t)|_{t=t_{0}}=U(x,t_{0})\,\,,\,\,\,\,\,\,\,\,-\infty<x<\infty
\end{array}
\right.
\end{eqnarray}
has solution in the following forms:
\begin{eqnarray}\label{sol-U}
a)\,\,\,\,\,\,\,\,\,\,\,\,\,\,\,\,\,\,\,\,\,\,\,\,\,\,\,\,\,\,\,\,\,\,\,\,
U(x,t)=\frac{\dot{r}(t)}{r(t)}x+\frac{r(t_{0})}{\mu(t)r(t)}V\left(\eta(x,t),\tau(t)\right),\,
\end{eqnarray}
where
\begin{equation}\label{eta-tau}
\eta(x,t)=\frac{r(t_{0})}{r(t)}x;\,\,\,\,
\,\,\,\,\,\tau(t)=r^{2}(t_{0})\int^{t}\frac{d\xi}{\mu(\xi)r^{2}(\xi)}\,\,,\,\,\tau(t_{0})=0,
\end{equation}
and the function $V(\eta,\tau)$ satisfies the IVP for the standard BE
\begin{eqnarray}\label{HBE-ic1}
\left\{
\begin{array}{ll}
  V_{\tau}+V V_{\eta}=\frac{1}{2}V_{\eta\eta},\,\\
V(\eta,0)=\mu(t_{0})U(\eta,t_{0}).
\end{array}
\right.
\end{eqnarray}

\begin{eqnarray}\label{sol-U2}
b)\,\,\,\,\,\,\,\,\,\,\,\,\,\,\,\,\,\,\,\,\,\,\,\,\,\,\,\,\,\,\,\,\,\,\,\,
U(x,t)=\frac{\dot{r}(t)}{r(t)}x-
\frac{r(t_{0})}{\mu(t)r(t)}\frac{\varphi_{\eta}(\eta(x,t),\tau(t))}{\varphi(\eta(x,t),\tau(t))},
\end{eqnarray}
where $\eta,$ $\tau$ are as defined in part (a), and $\varphi(\eta,\tau)$ satisfies the IVP for the heat equation
 \begin{eqnarray}\label{heat-ic1}
\left\{
\begin{array}{ll}
   \varphi_{\tau}=\frac{1}{2}\varphi_{\eta\eta}\,,\\
\varphi(\eta,0)=\exp\left[-\int^{\eta}\mu(t_{0})U(\xi,t_{0})d\xi\right].
\end{array}
\right.
\end{eqnarray}
\end{proposition}

\textbf{Proof:} a) Using the Ansatz  $ \,\,U(x,t)=[\mu(t)]^{-1}[\rho(t)x+s(t)V(s(t)x, \tau(t))],\,\,$
it is easy to show that, if the auxiliary functions satisfy the nonlinear system of ordinary differential equations
\begin{eqnarray}\label{aux}
\dot{\rho}+\frac{\rho^{2}}{\mu(t)}+\mu(t)\omega^{2}(t)=0,\,\,\,\,\,\,\rho(t_{0})=0,\\
\dot{\tau}-\frac{s^{2}}{\mu(t)}=0,\,\,\,\,\,\,\tau(t_{0})=0,\nonumber\\
\dot{s}+\frac{\rho(t)}{\mu(t)}s=0,\,\,\,\,\,s(t_{0})=1,\nonumber
\end{eqnarray}
 then the IVP (\ref{burgers-ic1}) for the NHBE transforms to the IVP (\ref{HBE-ic1}) for the standard BE. Also, noticing that  Eq.(\ref{aux}) is a nonlinear Riccati equation, the system is easily solved, and we obtain
  the  functions
\begin{equation}\label{a-tau-s} \rho(t)=\mu(t)\frac{\dot{r}(t)}{r(t)}\,;\,\,\,\,\,\,\,\,\,\,\,\,\,\,\tau(t)=
r^{2}(t_{0})\int^{t}\frac{d\xi}{\mu(\xi)r^{2}(\xi)}\,,\,\,\,\tau(t_{0})=0\,;
\,\,\,\,\,\,\,\,\,\,\,\,\,\,s(t)=\frac{r(t_{0})}{r(t)},
\end{equation}
 which  substituted back in the Ansatz give the result (\ref{sol-U}).
 Thus, solution  of the NHBE (\ref{burgers-ic1}) is explicitly obtained
 in terms of solution $V(\eta,\tau)$ to the BE (\ref{HBE-ic1}) and solution $r(t)$ of the IVP  for the linear ODE (\ref{clas-eq1}).  Part b) of the proposition, follows directly from the Cole-Hopf transformation $V=-\varphi_{\eta}/\varphi,$ which reduces the IVP (\ref{HBE-ic1}) for the BE to the  IVP (\ref{heat-ic1}) for the usual heat equation.
 $\Box$

 As well known, the IVP (\ref{heat-ic1}) for the heat equation has solution
 \begin{eqnarray*}
 \varphi(\eta,\tau)=\frac{1}{\sqrt{2\pi\tau}}
 \int_{-\infty}^{\infty}\exp\left[-\frac{(\eta-\xi)^2}{2\tau}\right]\varphi(\xi,0)d\xi,
 \end{eqnarray*}
and Cole-Hopf transformation $V=-\varphi_{\eta}/\varphi,$ leads to  solution of the IVP (\ref{HBE-ic1}) for the BE
\begin{eqnarray*}
V(\eta,\tau)=\frac{\int_{-\infty}^{\infty}
\left(\frac{\eta-\xi}{\tau}\right)\exp\left[-\left(\frac{(\eta-\xi)^2}{2\tau}+\int^{\xi}V(\xi ' ,0)d \xi ' \right)\right]d\xi}
{\int_{-\infty}^{\infty}
\exp\left[-\left(\frac{(\eta-\xi)^2}{2\tau}+\int^{\xi}V(\xi ' ,0)d\xi ' \right)\right]d\xi}.
\end{eqnarray*}
Therefore, using  the above proposition, one can find formal solution of the IVP (\ref{burgers-ic1}) for the NHBE  in terms of solution $r(t)$ of the linear ODE (\ref{clas-eq1}), that is
\begin{eqnarray}\label{gen-sol}
U(x,t)=\frac{\dot{r}(t)}{r(t)}x+\left[\frac{r(t_{0})}{\mu(t)r(t)}\right]\frac{\int_{-\infty}^{\infty}
\left(\frac{\frac{r(t_{0})}{r(t)}x-\xi}{\tau(t)}\right)
\exp\left[-\left(\frac{(\frac{r(t_{0})}{r(t)}x-\xi)^2}{2\tau(t)}+\int^{\xi}\mu(t_{0})U(\xi ' ,t_{0})d \xi ' \right)\right]d\xi}
{\int_{-\infty}^{\infty}
\exp\left[-\left(\frac{(\frac{r(t_{0})}{r(t)}x-\xi)^2}{2\tau(t)}+\int^{\xi}\mu(t_{0})U(\xi ' ,t_{0})d\xi ' \right)\right]d\xi},
\end{eqnarray}
where $\tau(t)$ is as defined in (\ref{eta-tau}), and the time interval on which the solution exists depends on the properties of the auxiliary functions. Since it is difficult to analyze  solution (\ref{gen-sol}) for an arbitrary initial condition, in what follows, we consider particular problems for which the NHBE (\ref{burgers-ic1}) subject to some localized initial profiles has exact solutions and one can observe explicitly their behavior.  As known, the standard BE (\ref{HBE-ic1}) has different type of solutions, such as  shock  solitary waves, similarity, \emph{N}-wave and rational function solutions. This suggests us to look  for corresponding type of solutions for the NHBE with variable coefficients, as follows.

 \textbf{ a. Shock solitary wave solutions.} The standard BE (\ref{HBE-ic1}) has shock solitary wave solution  \begin{equation}\label{sol-wave}
 V(\eta,\tau)=c-A\tanh\left[A\left( \eta-c\tau+c_{0}\right)\right],
 \end{equation}
 where $A,\, c,\, c_{0}$ are arbitrary real constants.
 Then, the  NHBE (\ref{burgers-ic1}) with initial condition
  \begin{eqnarray*}
U(x,t_{0})
=\frac{1}{\mu(t_{0})}\left(c-
A\tanh\left[A x+c_{0}\right]\right)
\end{eqnarray*}
has generalized shock solitary wave solution
 \begin{eqnarray}\label{U-tanh-g}
U(x,t)
=\frac{\dot{r}(t)}{r(t)}x+\frac{r(t_{0})}{\mu(t)r(t)}\left(c-
A\tanh\left[A\left( \frac{r(t_{0})}{r(t)}x-c\tau(t)+c_{0}\right)\right]\right),
\end{eqnarray}
where $r(t)$ is a solution of the IVP (\ref{clas-eq1}) and $\tau(t)$ is given in (\ref{eta-tau}).
In particular, when in (\ref{U-tanh-g}) one has $c=c_{0}=0,$  the NHBE (\ref{burgers-ic1}) has shock type static solution of the form
\begin{equation}\label{U-tanh}
U(x,t)=\frac{\dot{r}(t)}{r(t)}x-\frac{Ar(t_{0})}{\mu(t)r(t)}\tanh\left[\frac{Ar(t_{0})}{r(t)}x\right],
\end{equation}
corresponding to the initial condition $U(x,t_{0})=-(A/\mu(t_{0}))\tanh[A x]\,,\,-\infty<x<\infty.$

Solution (\ref{sol-wave}) of the standard BE (\ref{HBE-ic1}) is a localized wave of constant amplitude  moving with constant speed. However, the solution (\ref{U-tanh-g}) of the NHBE with variable coefficients is a generalized traveling wave of the form $U(x,t)=\tilde{u}(x,t)+A(t)\tanh[B(t)(x-C(t))],$ where the term $\tilde{u}(x,t)$ contributes to the wave amplitude, $A(t)$ is the shock amplitude, $B(t)$ is related with the steepness of the shock profile, $x=C(t)$ describes the motion of the "center" of the  profile, and $v=\dot{C}(t)$ is its velocity. Accordingly, for the wave solution (\ref{U-tanh-g}), one can see that the shock amplitude is proportional to $1/\mu(t)r(t),$ and steepness of the  profile is proportional to $1/r(t).$ Also, the position of the "center" of the wave profile is described by $x(t)=[r(t)/r(t_{0})](c\tau(t)-c_{0}),$ where its velocity can be easily found using that $v(t)=\dot{x}(t).$

 \emph{Multi-shock solitary wave solutions}. Since the standard BE has multi-shock solitary wave solutions, \cite{Whitham, Wang}, it is natural to ask for multi-shock solitary type solutions for the NHBE with variable coefficients. Here, we outline the procedure, and give formal results. Clearly, the heat equation (\ref{heat-ic1}), has simple solutions of the form
\begin{eqnarray*}
\varphi_{i}(\eta,\tau)=\exp\left[p_{i}(\eta,\tau)\right],\,\,\,\,\,\,
p_{i}(\eta,\tau)=-a_{i}\eta+\frac{a_{i}^{2}}{2}\tau+p_{i}^{0},\,\,\,\,\,\,a_{i},p_{i}^{0}\in \mathbb{R},
\end{eqnarray*}
and their linear superposition
\begin{eqnarray*}
\varphi(\eta,\tau)=\exp[p_{1}(\eta,\tau)]+\exp[p_{2}(\eta,\tau)]+...+\exp[p_{k}(\eta,\tau)],\,\,\,\,\,\,i=1,2,...,k,
\end{eqnarray*}
is also a solution.  By the Cole-Hopf transform $\,V=-\varphi_{\eta}/\varphi,\,$ it follows that the BE  (\ref{HBE-ic1}), has corresponding solutions of the form
\begin{eqnarray*}
V(\eta,\tau)=\frac{a_{1}\exp[p_{1}(\eta,\tau)]+a_{2}\exp[p_{2}(\eta,\tau)]+...+a_{k}\exp[p_{k}(\eta,\tau)]}
{\exp[p_{1}(\eta,\tau)]+\exp[p_{2}(\eta,\tau)]+...+\exp[p_{k}(\eta,\tau)]}.
\end{eqnarray*}
Therefore, using Proposition 2.1, one obtains that the NHBE (\ref{burgers-ic1}),
has generalized multi-shock solitary wave solutions given by
\begin{eqnarray*}
U(x,t)=\frac{\dot{r}(t)}{r(t)}x+\frac{r(t_{0})}{\mu(t)r(t)}\left[ \frac{a_{1}\exp[p_{1}(x,t)]+a_{2}\exp[p_{2}(x,t)]+...+a_{k}\exp[p_{k}(x,t)]}
{\exp[p_{1}(x,t)]+\exp[p_{2}(x,t)]+...+\exp[p_{k}(x,t)]}\right],
\end{eqnarray*}
where
$$p_{i}(x,t)=-a_{i}\frac{r(t_{0})}{r(t)}x+\frac{a_{i}^{2}}{2}\tau(t)+p_{i}^{0},\,\,\,\,\,\,\,\,\,\,\,\,i=1,2,...,k,$$
$r(t)$ is solution of  (\ref{clas-eq1}), and $\tau(t)$ is given in (\ref{eta-tau}).
For $k=2$ and proper choice of constants, one obtains one-shock solitary wave. When $k>2$ one expects formation of multi-shock solitary type solutions.  Indeed, this is the case, and illustrative examples are given in Sec.3.

 \textbf{b. Triangular wave solution.} The BE (\ref{HBE-ic1}) has triangular wave (similarity) solution,
 \begin{eqnarray}\label{h-sim-sol}
 V(\eta,\tau)=\frac{1}{\sqrt{2\pi \tau}}\left(\frac{(e^{A}-1)\exp[-\eta^{2}/2\tau]}
 {1+\frac{1}{2}(e^{A}-1) erfc[\eta/\sqrt{2\tau}]}\right),\,\,\,\,\,\,
 \end{eqnarray}
 corresponding to initial condition $V(\eta,0)=A\delta(\eta),$ see \cite{Whitham}, where $A$ is a constant, $\delta(\eta)$ is the Dirac-delta distribution, and $ erfc[a]=
 (2/\sqrt{\pi})\int_{a}^{\infty}\exp[-\xi^2]d\xi$ .
  Then, the NHBE (\ref{burgers-ic1}) has generalized triangular wave solution of the form
  \begin{eqnarray}\label{sim-solu}
  U(x,t)
=\frac{\dot{r}(t)}{r(t)}x+\frac{r(t_{0})}{\mu(t)r(t)}\frac{1}{\sqrt{2\pi \tau(t)}}
\left(\frac{(e^{A}-1)\exp\left[-(\frac{r(t_{0})}{r(t)}x)^{2}/2\tau(t)\right]}
{1+\frac{1}{2}(e^{A}-1) erfc\left[\frac{r(t_{0})}{r(t)}x/\sqrt{2\tau(t)}\right]}\right),\,\,\,\,\,\,\,\,\,\tau(t)>0.
  \end{eqnarray}

 \textbf{c. \emph{N}-wave solution.} The heat equation  (\ref{heat-ic1}) has solution
 \begin{eqnarray*}
 \varphi(\eta,\tau)=1+\sqrt{\frac{a}{\tau}}\exp[-\eta^{2}/2\tau],\,\,\,\,\,\,\,\,\,\tau>0,
 \end{eqnarray*}
 which  behaves like delta distribution as $\tau\rightarrow 0,$ (a-positive constant).
 The corresponding \emph{N}-wave solution of  BE  (\ref{HBE-ic1}), see \cite{Whitham}, is
 \begin{eqnarray*}
 V(\eta,\tau)=\left(\frac{\eta}{\tau}\right)
 \frac{\sqrt{a/\tau}\,\exp[-\eta^{2}/2\tau]}{1+\sqrt{a/\tau}\,\exp[-\eta^{2}/2\tau]},\,\,\,\,\,\,\,\,\tau>0,
 \end{eqnarray*}
Therefore, generalized \emph{N}-wave solution  of the NHBE (\ref{burgers-ic1}) is of the form
\begin{eqnarray}\label{N-wave}
U(x,t)
=\frac{\dot{r}(t)}{r(t)}x+\frac{r^{2}(t_{0})x}{\mu(t)\tau(t)r^{2}(t)}
\left(\frac{\sqrt{a/\tau(t)}\,\exp[-(\frac{r(t_{0})}{r(t)}x)^{2}/2\tau(t)]}
{1+\sqrt{a/\tau(t)}\,\exp[-(\frac{r(t_{0})}{r(t)}x)^{2}/2\tau(t)]}\right),\,\,\,\,\,\,\,\,\,\,\,\tau(t)>0.
\end{eqnarray}
Since the behavior of this solution at $\tau(t_{0})=0$ is rather complicated, as an initial profile one can consider a profile at any time $t>t_{0}.$

 \textbf{d. Rational function solutions.} Formal solution of the IVP for the heat equation (\ref{heat-ic1}) can be found also by applying the evolution operator to the initial condition, that is  $$\varphi(\eta,\tau)=\exp[(\tau/2)\partial^{2}_{\eta}]\varphi(\eta,0)
 =\sum_{k=0}^{\infty}\frac{1}{k!}(\frac{\tau}{2})^{k}\partial^{2k}_{\eta}\varphi(\eta,0).$$
  If the initial condition is $\,\varphi_{m}(\eta,0)=\eta^{m},$ then
   the solution of the heat problem (\ref{heat-ic1}) is $$\varphi_{m}(\eta,\tau)=\exp[(\tau/2)\partial^{2}_{\eta}]\eta^{m}=H_{m}(\eta,\tau/2),\,\,\,\,\,\,\,\,m=0,1,2,...,$$
  where $H_{m}(\eta,\tau/2)$ are Kamp\`{e} de Feriet polynomials, defined by
   $$H_{m}(\eta,\tau/2)=m!\sum_{k=0}^{[m/2]}\frac{(\tau/2)^k}{k!(m-2k)!}\eta^{m-2k},\,\,\,\,\,\,\,\,\,H_{m}(\eta,0)=\eta^{m},$$
with $[m/2]=m/2$ for even $m,$  and $[m/2]=(m-1)/2$ for odd $m,$ see \cite{Dattoli}. Using also  the relation $\partial _{\eta}H_{m}(\eta,\tau/2)=m H_{m-1}(\eta,\tau/2),$ it
   follows that, the BE (\ref{HBE-ic1})  has rational  solutions of the form
  \begin{equation}
  V_{m}(\eta,\tau)=-\frac{\partial}{\partial \eta}\left[\ln\varphi_{m}(\eta,\tau)\right]=-\frac{m H_{m-1}(\eta,\tau/2)}{H_{m}(\eta,\tau/2)},\,\,\,\,\,\,\,\,\,\,m=1,2,3,...,
  \end{equation}
or more generally
\begin{equation}
V_{k}(\eta,\tau)=-\frac{\partial}{\partial \eta}\left(\ln \sum_{m=0}^{k}a_{m}\varphi_{m}(\eta,\tau)\right)=-
\frac{\sum_{m=1}^{k}m a_{m}H_{m-1}(\eta,\tau/2)}{\sum_{m=0}^{k}a_{m}H_{m}(\eta,\tau/2)},\,\,\,\,k=1,2,3,...,
\end{equation}
where $a_{m}$ are arbitrary real constants. Therefore, we obtain that the variable coefficient NHBE (\ref{burgers-ic1}) with initial conditions
$U_{m}(x,t_{0})=-(m/\mu(t_{0})x),\,\,\,\,m=1,2,3,...,$ has rational solutions
\begin{equation}\label{um}
U_{m}(x,t)=\frac{\dot{r}(t)}{r(t)}x-\frac{m}{\mu(t)} (\frac{r(t_{0})}{r(t)})\left(\frac{H_{m-1}\left(\frac{r(t_{0})}{r(t)}x,\frac{1}{2}\tau(t)\right)}
{H_{m}\left(\frac{r(t_{0})}{r(t)}x,\frac{1}{2}\tau(t)\right)}\right),
\end{equation}
and with more general initial conditions
\begin{eqnarray*}
U_{k}(x,t_{0})=-\frac{1}{\mu(t_{0})}\left(\frac{\sum_{m=1}^{k}m a_{m}x^{m-1}}{\sum_{m=0}^{k}a_{m}x^{m}}\right),\,\,\,\,\,\,k=1,2,3,...
\end{eqnarray*}
 it has rational solutions of the form
\begin{equation}\label{gu}
U_{k}(x,t)=\frac{\dot{r}(t)}{r(t)}x-\frac{1}{\mu(t)} (\frac{r(t_{0})}{r(t)})\left(\frac{\sum_{m=1}^{k}m a_{m}H_{m-1}\left(\frac{r(t_{0})}{r(t)}x,\frac{1}{2}\tau(t)\right)}
{\sum_{m=0}^{k}a_{m}H_{m}\left(\frac{r(t_{0})}{r(t)}x,\frac{1}{2}\tau(t)\right)}\right).
\end{equation}
The  solutions (\ref{um}) and (\ref{gu}) can be written also in terms of  standard Hermite polynomials  $H_{m}(y)$ using that  \begin{equation}\label{KFP-st}
H_{m}(\eta,\tau/2)=
\frac{(\sqrt{\tau/2})^{m}}{i^{m}}H_{m}\left(\frac{i\eta}{2\sqrt{\tau/2}}\right),\,\,\,\,\,\,\,\,\,\,
H_{m}(\eta,0)=\eta^{m},\,\,\,\,\,\,m=0,1,2,...,
\end{equation}
 where $H_{m}(y)\,$ are   defined by
$\,\,\exp[2y\xi-\xi^{2}]=\sum_{m=0}^{\infty}(\xi^m/m!)H_{m}(y).$
Thus,  the points where Kamp\`{e} de Feriet  polynomials vanish can be found in terms of the well known zeros of the Hermite polynomials. For this, we denote by $\,y_{m}^{(l)},\,\,\,l=1,2,...,m\,$ the zeros of the Hermite polynomial $H_{m}(y),$ so that for each fixed $m$, one has $H_{m}(y_{m}^{(l)})=0$ for all $l=1,2,...,m.$ From relation (\ref{KFP-st}) it follows that
\begin{eqnarray}\label{herm-zeros}
H_{m}(\eta,\tau/2)=0 \,\,\,\,\,\,\,\,\,\,  \Longleftrightarrow \,\,\,\,\,\,\,\,\,\,  \eta=-i\,2y_{m}^{(l)}\sqrt{\tau/2},\,\,\,\,\,\,\,\,\,l=1,2,...,m.
\end{eqnarray}
Thus,  $U_{m}(x,t)$ given by (\ref{um}) has singularities  at points where $H_{m}\left((r(t_{0})/r(t))x,\tau(t)/2\right)=0,$ and according to (\ref{herm-zeros}), the motion of these pole singularities  is described by
\begin{eqnarray}\label{poles}
x_{m}^{(l)}(t)=-i\,2 y_{m}^{(l)}\frac{r(t)}{r(t_{0})}\sqrt{\frac{\tau(t)}{2}},\,\,\,\,\,\,\,\,\,\, l=1,2,...,m.
\end{eqnarray}
We note that, for a real-valued solution $r(t)$  and $\tau(t)>0,\,\,\,\,t\in I,$  the solution $U_{m}(x,t)$ does not have moving singularities on the real line.  It may have real singularity only at $x=0.$  On the other hand, for some special choice of the coefficients $a_{m},$ the solutions $U_{k}(x,t)$ of the form (\ref{gu}) may have  singularities moving on the real line. Illustrative examples are given in next section.

 \section{Exactly solvable Burgers models}

 \subsection{Forced Burgers equation with constant coefficients }

  Burgers equation with constant coefficients and a forcing term linear in the space variable $x$
   \begin{eqnarray}\label{st-bur}
   U_{t}+U U_{x}=\frac{1}{2}U_{xx}-\omega_{0}^{2}x,
   \end{eqnarray}
   is a known integrable model and one can see for example \cite{Eule,Axel-2}.
    For this model, according to Proposition 2.1 one has  $\mu(t)=1,$ $\gamma=0$, and $\omega^{2}(t)=\omega^{2}_{0}\,-$ real constant, so that the corresponding IVP for the second order linear ODE is
   \begin{eqnarray}\label{st-ode}
   \ddot{r}(t)+\omega_{0}^{2}r(t)=0, \,\,\,\,r(0)=r_{0}\neq 0,\,\,\dot{r}(0)=0.
   \end{eqnarray}
 When, $\omega^{2}_{0}=0,$ one has $r(t)=r_{0},$ $\eta=x,$ $\tau=t,$ and formula (\ref{sol-U}) gives $U(x,t)=V(x,t),$ which is a solution of the standard Burgers equation, as expected. Using the approach in previous section, we will recall some particular solutions for the cases when  $\omega^{2}_{0}>0$ and $\omega^{2}_{0}<0.$

\subsubsection{Case $\omega^{2}_{0}>0.$} In that case the IVP (\ref{st-ode})
has oscillating solution $r(t)=r_{0}\cos(\omega_{0}t)$ and the auxiliary function is
$\tau(t)=\tan(\omega_{0}t)/\omega_{0}.$ From Proposition 2.1, it follows that the forced BE (\ref{st-bur}) has solutions
\begin{eqnarray}\label{f-w1}
U(x,t)=-\omega_{0}\tan(\omega_{0}t)x+\sec(\omega_{0}t)V(\sec(\omega_{0}t)x,\tan(\omega_{0}t)/\omega_{0}),
\end{eqnarray}
where $V(\eta,\tau)$ is a solution of the standard Burgers equation.  In what follows, using the discussion in previous section, we will write explicitly some special solutions of BE (\ref{st-bur}), and  note that, in the limit case $\omega_{0}\rightarrow 0,$  these solutions $U(x,t)$ approach the solutions $V(x,t)$ of the standard BE.

\textbf{a.} Forced Burgers equation (\ref{st-bur}) with initial condition
 $U(x,0)=-A\tanh(A x)\,,\,\,-\infty<x<\infty,$
 has shock type static wave solution
\begin{eqnarray}\label{con-burg}
U(x,t)=-\omega_{0}\tan(\omega_{0}t)x-A\sec(\omega_{0}t)\tanh\left(A\sec(\omega_{0}t)x\right),
\end{eqnarray}
and with initial condition
 $U(x,0)=c-A\tanh[A x+c_{0}]\,\,,\,\-\infty<x<\infty,$
it has shock solitary type solution of the form
 \begin{eqnarray}\label{c-BE-2}
U(x,t)
=-\omega_{0}\tan(\omega_{0}t)x+\sec(\omega_{0}t)\left(c-A\tanh\left[A\left( \sec(\omega_{0}t) x-\frac{c}{\omega_{0}}\tan(\omega_{0}t)\right)+c_{0}\right]\right).
\end{eqnarray}

\textbf{b.} Forced BE (\ref{st-bur}) has triangular wave solution
\begin{eqnarray*}\label{sim-BE-st}
U
=-\omega_{0}\tan(\omega_{0}t)x+\sec(\omega_{0}t)\sqrt{\frac{\omega_{0}}{2\pi\tan(\omega_{0}t)}}
\left(\frac{(e^{A}-1)\exp\left[-
\frac{(\sec(\omega_{0}t)x)^{2}}{2\tan(\omega_{0}t)/\omega_{0}}\right]}
{1+\frac{1}{2}(e^{A}-1)\, erfc\left[\frac{\sec(\omega_{0}t)x}{ \sqrt{2\tan(\omega_{0}t)/\omega_{0}}}\right]}\right),
\end{eqnarray*}
for $t\in(0,\pi/2\omega_{0}).$ In general, triangular wave exists on time interval where $\tau(t)=\tan(\omega_{0}t)/\omega_{0}>0.$

\textbf{c.} \emph{N}-wave solution can also be considered  for BE (\ref{st-bur}) using formula (\ref{N-wave}) on time interval where $\tau(t)>0.$

\textbf{d.} Forced BE (\ref{st-bur}) subject to initial conditions $
U_{m}(x,0)=-m/x\,\,,\,m=1,2,3,...,$
has rational function solutions of the form
\begin{eqnarray*}
U_{m}(x,t)=-\omega_{0}\tan(\omega_{0}t)x-\sec(\omega_{0}t)
\left[\frac{m H_{m-1}\left(\sec(\omega_{0}t)x,\tan(\omega_{0}t)/2\omega_{0}\right)}
{H_{m}\left(\sec(\omega_{0}t)x,\tan(\omega_{0}t)/2\omega_{0}\right)}\right],
\end{eqnarray*}
with moving pole singularities
$x_{m}^{(l)}(t)=-iy_{m}^{l}\cos(\omega_{0}t)\sqrt{2\tan(\omega_{0}t)/\omega_{0}},
\,\,\,\,\,l=1,2,...,m.$

\subsubsection{Case $\omega^{2}_{0}<0.$} When $\omega^{2}_{0}=-\tilde{\omega}^2,$  $\tilde{\omega}>0,$ the BE (\ref{st-bur}) becomes
\begin{eqnarray}\label{st-bur-2}
   U_{t}+U U_{x}=\frac{1}{2}U_{xx}+\tilde{\omega}^{2}x,
   \end{eqnarray}
and the corresponding ODE (\ref{st-ode}) has solution $r(t)=r_{0}\cosh(\tilde{\omega}t),$ where the auxiliary function is $\tau(t)=\tanh(\tilde{\omega}t)/\tilde{\omega}.$ Therefore, solutions of BE (\ref{st-bur-2}) are of the form
\begin{eqnarray}\label{solu-1}
U(x,t)=\tilde{\omega}\tanh(\tilde{\omega}t)x+\frac{1}{\cosh(\tilde{\omega}t)}
V\left(\frac{x}{\cosh(\tilde{\omega}t)},\frac{\tanh(\tilde{\omega}t)}{\tilde{\omega}}\right),
\end{eqnarray}
 where $V(\eta,\tau)$ is a solution of the standard BE. As in the previous case, we see that in the limit $\tilde{\omega}\rightarrow 0$, the following particular exact solutions $U(x,t)$  approach the corresponding solution $V(x,t)$ of the standard BE.

 \textbf{a.} The forced BE (\ref{st-bur-2}) with initial condition $U(x,0)=c-A\tanh[Ax],\,\,-\infty<x<\infty$  has one-shock solitary type solution
\begin{eqnarray}\label{f-solitary}
U(x,t)=\tilde{\omega}\tanh(\tilde{\omega}t)x
 +\frac{1}{\cosh(\tilde{\omega}t)}
 \left(c-A\tanh\left[\frac{A}{\cosh(\tilde{\omega}t)}(x-c\frac{\sinh(\tilde{\omega}t)}{\tilde{\omega}})\right]\right),
\end{eqnarray}
which amplitude depends on time, the center of the wave profile moves according to $x(t)=c\sinh(\tilde{\omega}t)/\tilde{\omega},$ and its velocity is $v(t)=c\cosh(\tilde{\omega}t).$

In general, BE (\ref{st-bur-2}) has multi-shock solitary type solutions of the form
\begin{eqnarray}\label{f-multi}
U(x,t)=\tilde{\omega}\tanh(\tilde{\omega}t)x+\frac{1}{\cosh(\tilde{\omega}t)}
 \left[ \frac{a_{1}\exp[p_{1}(x,t)]+a_{2}\exp[p_{2}(x,t)]+...+a_{k}\exp[p_{k}(x,t)]}
{\exp[p_{1}(x,t)]+\exp[p_{2}(x,t)]+...+\exp[p_{k}(x,t)]}\right],
\end{eqnarray}
where $$p_{i}(x,t)=-\frac{a_{i}}{\cosh(\tilde{\omega}t)}x
+\frac{a_{i}^{2}}{2\tilde{\omega}}\tanh(\tilde{\omega}t)+p_{i}^{0},\,\,\,\,\,i=1,2,....$$

\textbf{b.} Triangular wave solution of BE (\ref{st-bur-2}) is
\begin{eqnarray*}
U(x,t)=\tilde{\omega}\tanh(\tilde{\omega}t)x+
\frac{1}{\cosh(\tilde{\omega}t)} \sqrt{\frac{\tilde{\omega}}{2\pi \tanh(\tilde{\omega}t)}}
\left(\frac{(e^{A}-1)\exp\left[-(\frac{x}
{\cosh(\tilde{\omega}t)})^{2}/(\frac{2\tanh(\tilde{\omega}t)}{\tilde{\omega}})\right]}
{1+\frac{1}{2}(e^{A}-1)erfc\left[
(\frac{x}{\cosh(\tilde{\omega}t)})/\sqrt{\frac{2\tanh(\tilde{\omega}t)}{\tilde{\omega}}}\right]}
\right),
\end{eqnarray*}
which corresponds to Dirac-delta initial distribution.

\textbf{c.} \emph{N}-wave solution of BE (\ref{st-bur-2}) is of the form
\begin{eqnarray}\label{f-c-N-wave}
U(x,t)=\tilde{\omega}\tanh(\tilde{\omega}t)x+\frac{\tilde{\omega} x}{\sinh(\tilde{\omega}t)\cosh(\tilde{\omega}t)}
\left[\frac{\sqrt{\frac{a\tilde{\omega}}{\tanh(\tilde{\omega}t)}}
\exp\left[-(\frac{x}
{\cosh(\tilde{\omega}t)})^{2}/2(\frac{\tanh(\tilde{\omega}t)}{\tilde{\omega}})\right]}{1+\sqrt{\frac{
a\tilde{\omega}}{\tanh(\tilde{\omega}t)}}\exp\left[-(\frac{x}
{\cosh(\tilde{\omega}t)})^{2}/2(\frac{\tanh(\tilde{\omega}t)}{\tilde{\omega}})\right]}\right].
\end{eqnarray}

\textbf{d.} Rational function solutions of BE (\ref{st-bur-2}) subject to initial conditions $\,
U_{m}(x,0)=-m /x\,\,,\,m=1,2,3,...,$
are of the form
\begin{eqnarray}\label{f-st-rat}
U_{m}(x,t)=\tilde{\omega}\tanh(\tilde{\omega}t)x+\frac{1}{\cosh(\tilde{\omega}t)}
\left(\frac{mH_{m-1}\left( x/\cosh(\tilde{\omega}t),\tanh(\tilde{\omega}t)/2\tilde{\omega}\right)}
{H_{m}\left( x/\cosh(\tilde{\omega}t),\tanh(\tilde{\omega}t)/2\tilde{\omega}\right)  }\right).
\end{eqnarray}
For each fixed $m,$ the rational solution (\ref{f-st-rat}) has pole singularities whose motion in the complex plane is described by
$$x_{m}^{(l)}(t)=-iy_{m}^{(l)}\cosh(\tilde{\omega}t)\sqrt{2 (\frac{\tanh(\tilde{\omega}t)}{\tilde{\omega}})},
 \,\,\,\,\,\,\,\,t>0,\,\,\,\,\,\,l=1,2,...,m,$$
 and as $\,\tilde{\omega}\rightarrow 0\,$ they approach the well known poles $\,\,x_{m}^{(l)}(t)=-iy_{m}^{(l)}\sqrt{2t}\,\,$ of  the standard BE.

 Finally, we note that for the above particular solutions, one has $\lim_{t\rightarrow\infty}U(x,t)=\tilde{\omega}x.$ Similar result was obtained in \cite{Ding} and \cite{Yadav}, where
 the long-time asymptotics for solutions of the BE (\ref{st-bur-2}) were discussed according to the properties of the initial profile.

\subsection{Forced Burgers equation with constant damping and exponentially decaying diffusion coefficient}

In this part, we consider exactly solvable forced Burgers equations of the form
\begin{eqnarray}\label{cal-kanai}
 U_{t}+\gamma U+U U_{x}=\frac{1}{2}e^{-\gamma t}U_{xx}-\omega_{0}^{2}x\,\,,\,\,\,\,\,\,\,\,\,-\infty<x<\infty,\,\,\,\,t>0,
\end{eqnarray}
 with constant damping $\Gamma(t)=\gamma>0,$ exponentially decaying diffusion coefficient $D(t)=e^{-\gamma t}/2,$  $\omega^{2}(t)=\omega_{0}^{2}>0$ and $\mu(t)=e^{\gamma t}.$ The corresponding IVP for the linear ODE is then
\begin{equation}\label{clas-par}
\ddot{r}+\gamma \dot{r}+\omega_{0}^{2}r=0,\,\,\,\,\,r(0)=r_{0}\neq
0,\,\,\,\dot{r}(0)=0,
\end{equation}
  and it has three different type of solutions depending on  $\,\omega_{0}^{2}-(\gamma^{2}/4).\,$ In what follows, for each case we discuss separately the related variable coefficient Burgers equations (\ref{cal-kanai}).

 \subsubsection{ Critical damping case.}

 If $\omega_{0}^{2}-(\gamma^{2}/4)=0$, the IVP
  (\ref{clas-par}) for the linear ODE, has solution
\begin{equation}
r_{1}(t)=r_{1}(0)e^{-\frac{\gamma t}{2}}(1+\frac{\gamma}{2}t),
\end{equation}
 and the auxiliary function is
 $\,\tau_{1}(t)=t/(1+\gamma t/2).\,$
 Therefore,  the BE (\ref{cal-kanai}) has  solutions of the form
 \begin{eqnarray}\label{sol-U-1}
U(x,t)=-(\frac{\gamma}{2})^{2}\left(\frac{t}{1+\frac{\gamma}{2}t}x\right)+
\left(\frac{e^{-\gamma t/2}}{1+\frac{\gamma}{2}t}\right)V\left(\frac{e^{\gamma t/2}x}{1+\frac{\gamma}{2}t},\frac{t}{1+\frac{\gamma}{2}t}\right),
\end{eqnarray}
 where $V(\eta,\tau)$ satisfies the standard BE.

   Clearly, when $\omega_{0}\rightarrow 0,$ one has also $\gamma \rightarrow 0,$ and  in that case we can see that the following  particular solutions $U(x,t)$  approach  solutions $V(x,t)$ of the standard BE.

\begin{figure}[h!]
{\includegraphics[width=7cm,height=6cm]{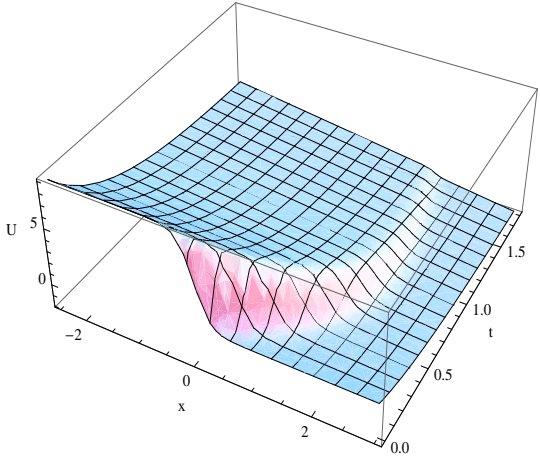}}
\caption{ Critical damping case.  One-shock solitary type solution $U(x,t)$ given by (\ref{crit-soli}), $\,\gamma=2,$ $A=4, c=5, c_{0}=0.$ }
\label{ivpb1}
\end{figure}

\begin{figure}[h!]
{\includegraphics[width=7cm,height=6cm]{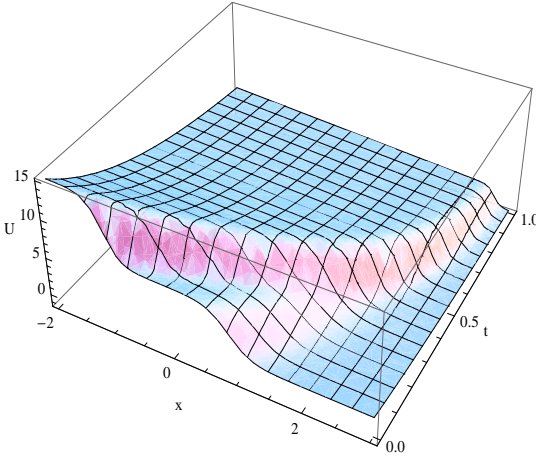}}
\caption{Critical damping case.  Fusion of two-shock solitary type solution $U(x,t).$ }
\label{crit-multi}
\end{figure}

\textbf{a.} BE (\ref{cal-kanai}) with initial condition $U(x,0)=c-A\tanh\left[Ax\right]$
has shock solitary  type solution
\begin{eqnarray}\label{crit-soli}
U(x,t)=-(\frac{\gamma}{2})^{2}\left(\frac{t}{1+\frac{\gamma}{2}t}\right)x+
\left(\frac{e^{-\gamma t/2}}{1+\frac{\gamma}{2}t}\right)
\left(c-A\tanh\left[A\left( \frac{e^{\gamma t/2}}{1+\frac{\gamma}{2}t}(x-cte^{-\gamma t/2})\right)\right]\right),
\end{eqnarray}
 which  shock amplitude decays with time eventually going to zero,  and its  "center" moves with velocity $v(t)=c(1-\gamma t/2)e^{-\gamma t/2},\,$ see Fig.\ref{ivpb1}.
Multi-shock solitary type solutions of the BE (\ref{cal-kanai}), can be found from the general solution
\begin{eqnarray}\label{crit-multi-sh}
U(x,t)= -(\frac{\gamma}{2})^{2}\left(\frac{t}{1+\frac{\gamma}{2}t}\right)x+
\left(\frac{e^{-\gamma t/2}}{1+\frac{\gamma}{2}t}\right)
\left[ \frac{a_{1}\exp[p_{1}(x,t)]+...+a_{k}\exp[p_{k}(x,t)]}
{\exp[p_{1}(x,t)]+...+\exp[p_{k}(x,t)]}\right],
\end{eqnarray}
where
$$p_{i}(x,t)=-a_{i}\left(\frac{e^{\gamma t/2}}{1+\frac{\gamma}{2} t}
\right)x+\frac{a_{i}^{2}}{2}\left(\frac{t}{1+\frac{\gamma}{2}t}\right)+p_{i}^{0},\,\,\,\,\,\,\,\,\,\,\,\,i=1,2,...,k,$$
and $a_{i},p_{i}^{0}$ are real constants. In Fig.\ref{crit-multi} we plot two-shock solitary type wave solution $U(x,t),$ with special choices in (\ref{crit-multi-sh}), $k=3, \gamma=2,$ $a_{1}=1,$ $p_{1}^{0}=0,$ $a_{2}=7,$ $p_{2}^{0}=5,$ $a_{3}=15,$ $p_{3}^{0}=-4.$ We observe fusion of two-shock solitary wave, which shock contribution eventually goes to zero.

\textbf{b.} Using again the results in Sec.2, one obtains that the BE (\ref{cal-kanai}) has generalized triangular wave solution
\begin{eqnarray*}\label{sim-sol-crit}
U(x,t)
=-(\frac{\gamma}{2})^{2}\left(\frac{t}{(1+\frac{\gamma}{2}t)}\right)x
-\frac{ e^{-\frac{\gamma}{2}t}}{(1+\frac{\gamma}{2}t)}\sqrt{\frac{(1+\gamma t/2)}{2\pi t}}
\left(\frac{(e^{A}-1)\exp\left[-e^{\gamma t}x^2/2t(1+\gamma t/2)\right]}{1+\frac{1}{2}(e^{A}-1)
erfc\left[ e^{\gamma t/2} x /\sqrt{2t(1+\gamma t/2)}\right]}\right),
\end{eqnarray*}
corresponding to Dirac-delta initial profile.

\textbf{c.} BE (\ref{cal-kanai}) for the critical damping case has \emph{N}-wave solution of the form (\ref{N-wave}) with $r(t)=r_{1}(t)$ and $\tau(t)=\tau_{1}(t).$ Explicit form of this generalized \emph{N}-wave solution is
\begin{eqnarray*}\label{N-crit}
U(x,t)
=-(\frac{\gamma}{2})^{2}\left(\frac{t}{(1+\frac{\gamma}{2}t)}\right)x
+\frac{x}{t(1+\frac{\gamma}{2} t)}\left(\frac{\sqrt{\frac{a(1+\gamma t/2)}{t}}\exp\left[-e^{\gamma t}x^2/2t(1+\gamma t/2)\right]}{1+\sqrt{\frac{a(1+\gamma t/2)}{t}}\exp\left[-e^{\gamma t}x^2/2t(1+\gamma t/2)\right]}\right).
\end{eqnarray*}

\textbf{d.} Rational solutions of the BE (\ref{cal-kanai}) with initial conditions $U_{m}(x,0)=-m(1/x),\,\,m=1,2,...$ are of the form
\begin{eqnarray}\label{crit-ratio}
U_{m}(x,t)
=-(\frac{\gamma}{2})^{2}\left(\frac{t}{(1+\frac{\gamma}{2}t)}\right)x
-\frac{ e^{-\gamma t/2}}{(1+\frac{\gamma}{2}t)}
\left(\frac{m H_{m-1}\left(\frac{e^{\gamma t/2}}{1+\frac{\gamma}{2}t}x,
\,\frac{t}{2(1+\frac{\gamma}{2}t)}\right)}{H_{m}\left(\frac{e^{\gamma t/2}}
{1+\frac{\gamma}{2}t}x,\,\frac{t}{2(1+\frac{\gamma}{2}t)}\right)}\right).
\end{eqnarray}

 According to (\ref{poles}), for each $m,$ the motion of the pole singularities is described by
\begin{eqnarray}
x_{m}^{(l)}(t)=-i\,\sqrt{2}\,y_{m}^{(l)}e^{-\gamma t/2}\sqrt{t(1+\gamma t/2)}, \,\,\,l=1,2,....
\end{eqnarray}
 Since $\gamma>0$ and $t>0$, clearly $U_{m}(x,t)$ has no moving singularities on the real line.  On the other hand, forced BE (\ref{cal-kanai}) with  more general initial conditions
$$
U_{k}(x,0)=-(\sum_{m=1}^{k}m a_{m}x^{m-1})/\sum_{m=0}^{k}a_{m}x^{m},\,\,\,\,\,k=1,2,...,$$
has the following rational solutions
\begin{eqnarray}\label{g-sol}
U_{k}(x,t)
=-(\frac{\gamma}{2})^{2}\left(\frac{t}{(1+\frac{\gamma}{2}t)}\right)x
-\frac{ e^{-\gamma t/2}}{(1+\frac{\gamma}{2}t)}
\left(\frac{\sum_{m=1}^{k}m a_{m}H_{m-1}\left(\frac{e^{\gamma t/2}}{1+\frac{\gamma}{2}t}x,
\,\frac{t}{2(1+\frac{\gamma}{2}t)}\right)}{\sum_{m=0}^{k}a_{m}H_{m}\left(\frac{e^{\gamma t/2}}
{1+\frac{\gamma}{2}t}x,\,\frac{t}{2(1+\frac{\gamma}{2}t)}\right)}\right).
\end{eqnarray}
 
At the end of this section, we note that for the above particular solutions  one has  $\lim_{t\rightarrow \infty}U(x,t)=-(\gamma/2)x,$ so that in the long-time limit the system becomes stable with velocity proportional to the displacement. 

 \subsubsection{Under damping case.}

 If $\omega_{0}^{2}-(\gamma^{2}/4)>0,$ then
 the ODE (\ref{clas-par})
 has oscillating solution
\begin{equation}\label{xt-2}
r_{2}(t)=r_{2}(0)\frac{\omega_{0}}{\Omega}e^{-\frac{\gamma
t}{2}}\cos[\Omega t-\alpha],
\end{equation}
where  $\Omega=\sqrt{\omega_{0}^{2}-(\gamma^{2}/4)}>0,$ $\,\alpha=\tan^{-1}(\frac{\gamma}{2\Omega}),$ and the auxiliary function is
\begin{eqnarray}\label{tau-2}
      \tau_{2}(t)=\frac{ \Omega
}{\omega_{0}^{2}}\left(\tan[\Omega t-\alpha]+\frac{\gamma}{2\Omega}\right).
\end{eqnarray}
Then,  BE (\ref{cal-kanai})  has  solutions of the form
\begin{eqnarray}\label{sol-U-2}
U(x,t)&=&-\left(\frac{\gamma}{2}+\Omega\tan[\Omega t-\alpha]\right)x  \nonumber\\
& &+\frac{\Omega}{\omega_{0}}\frac{e^{-\gamma t/2}}{\cos[\Omega t-\alpha]}
V\left(  \frac{\Omega}{\omega_{0}}\frac{e^{\gamma t/2}x}{\cos[\Omega t-\alpha]},\frac{ \Omega
}{\omega_{0}^{2}}\left(\tan[\Omega t-\alpha]+\frac{\gamma}{2\Omega}\right)\right),
\end{eqnarray}
  where $V(\eta,\tau)$ satisfies  the standard BE.

   When $\gamma\rightarrow 0,$ one has $\Omega\rightarrow \omega_{0}>0$ and $\alpha\rightarrow 0.$ In that case, it is not difficult to see that the bellow given solutions of the forced BE (\ref{cal-kanai}) with variable coefficients  approach the corresponding solutions of the forced BE (\ref{st-bur}) with constant coefficients.

\textbf{a.}  BE (\ref{cal-kanai}) has shock type static  solution
\begin{eqnarray}\label{static-sol}
U(x,t)=-\left(\frac{\gamma}{2}+|\Omega|\tan[\Omega t-\alpha]\right)x-
\frac{A\Omega}{\omega_{0}}\frac{e^{-\gamma t/2}}{\cos[\Omega t-\alpha]}\tanh\left[\frac{A\Omega}{\omega_{0}}\frac{e^{\gamma t/2}x}{\cos[\Omega t-\alpha]}\right],
\end{eqnarray}
 and it has
 shock solitary type solution in the form
\begin{eqnarray}\label{soli-sol}
U(x,t)=-\left(\frac{\gamma}{2}+\Omega \tan[\Omega t-\alpha]\right)x\,\,\,\,\,\,\,\,\,\,\,\,\,\,  \nonumber\\
+\frac{\Omega}{\omega_{0}}\frac{e^{-\gamma t/2}}{\cos[\Omega t-\alpha]}\left(c-A\tanh\left[A\left(\frac{\Omega}{\omega_{0}}\frac{e^{\gamma t/2}x}{\cos[\Omega t-\alpha]}-c\frac{ \Omega
}{\omega_{0}^{2}}\left(\tan[\Omega t-\alpha]+\frac{\gamma}{2\Omega}\right)\right)\right]\right).
\end{eqnarray}
  Note that the solitary wave is broken by shock discontinuities which appear periodically at finite times.
Multi-shock solitary waves can be obtained using the solution
\begin{eqnarray*}
U(x,t)=\frac{\dot{r}_{2}(t)}{r_{2}(t)}x+e^{-\gamma t}\frac{r_{2}(t_{0})}{r_{2}(t)}\left[ \frac{a_{1}\exp[p_{1}(x,t)]+a_{2}\exp[p_{2}(x,t)]+...+a_{k}\exp[p_{k}(x,t)]}
{\exp[p_{1}(x,t)]+\exp[p_{2}(x,t)]+...+\exp[p_{k}(x,t)]}\right],
\end{eqnarray*}
where
$$p_{i}(x,t)=-a_{i}\frac{r_{2}(t_{0})}{r_{2}(t)}x+\frac{a_{i}^{2}}{2}\tau_{2}(t)+p_{i}^{0},\,\,\,\,\,\,\,\,\,\,\,\,i=1,2,...,k,$$
$r_{2}(t)$ is given by  (\ref{xt-2}), and $\tau_{2}(t)$ is given by (\ref{tau-2}).

\textbf{b.}  BE (\ref{cal-kanai}) has generalized triangular wave solution of the form
(\ref{sim-solu}) with $r(t)=r_{2}(t)$ and $\tau(t)=\tau_{2}(t)$ on a time  interval where $\tau_{2}(t)>0.$

\textbf{c.} Generalized \emph{N}-wave solution of  BE (\ref{cal-kanai}) for the under damping case can be found using formula (\ref{N-wave}) on a time  interval where $\tau_{2}(t)>0.$

\textbf{d.} BE (\ref{cal-kanai}) with initial conditions
$$
U_{k}(x,0)=-(\sum_{m=1}^{k}m a_{m}x^{m-1})/\sum_{m=0}^{k}a_{m}x^{m},\,\,\,\,\,k=1,2,...,$$
has  rational function solutions
\begin{equation}\label{gu-33}
U_{k}(x,t)=\frac{\dot{r}_{2}(t)}{r_{2}(t)}x-e^{-\gamma t} (\frac{r_{2}(0)}{r_{2}(t)})\left(\frac{\sum_{m=1}^{k}m a_{m}H_{m-1}\left(\frac{r_{2}(0)}{r_{2}(t)}x,\frac{1}{2}\tau_{2}(t)\right)}
{\sum_{m=0}^{k}a_{m}H_{m}\left(\frac{r_{2}(0)}{r_{2}(t)}x,\frac{1}{2}\tau_{2}(t)\right)}\right).
\end{equation}

\subsubsection{ Over damping case.} When $\omega_{0}^{2}-(\gamma^{2}/4)<0,$
 the  IVP (\ref{clas-par}) has solution
\begin{eqnarray}\label{x3}
r_{3}(t)=r_{3}(0)\frac{\omega_{0}}{\Omega'} e^{-\frac{\gamma t}{2}}
\sinh[\Omega' t+\beta],
\end{eqnarray}
where $\Omega'=\sqrt{|\omega_{0}^{2}-(\gamma^{2}/4)|},$ and
$\beta=\coth^{-1}(\frac{\gamma}{2\Omega'}).$ Then,
\begin{eqnarray}\label{a3,tau3}
\tau_{3}(t)=-\frac{\Omega'}{\omega_{0}^{2}}\left(\coth[\Omega' t+\beta]-\frac{\gamma}{2\Omega'}\right),
\end{eqnarray}
and thus  BE (\ref{cal-kanai})  has  solutions of the form
\begin{eqnarray}\label{sol-U-3}
U(x,t)&=&\left(-\frac{\gamma}{2}+\Omega'\coth[\Omega't+\beta]\right)x \nonumber\\
& &+\left(\frac{\Omega'e^{-\gamma t/2}}{\omega_{0}\sinh[\Omega't+\beta]}\right)\,V\left(\frac{\Omega'e^{\gamma t/2}x}{\omega_{0}\sinh[\Omega't+\beta]},-\frac{\Omega'}{\omega_{0}^{2}}\left(\coth[\Omega't+\beta]-\frac{\gamma}{2\Omega'})   \right)\right),
\end{eqnarray}
where $V(\eta,\tau)$ satisfies the standard BE (\ref{HBE-ic1}).

\textbf{a.} Shock and multi-shock solitary type solutions for the forced BE (\ref{cal-kanai}) can be obtained from the general expression
\begin{eqnarray}\label{multi-und}
U(x,t)=\frac{\dot{r}_{3}(t)}{r_{3}(t)}x+e^{-\gamma t}\frac{r_{3}(t_{0})}{r_{3}(t)}\left[ \frac{a_{1}\exp[p_{1}(x,t)]+a_{2}\exp[p_{2}(x,t)]+...+a_{k}\exp[p_{k}(x,t)]}
{\exp[p_{1}(x,t)]+\exp[p_{2}(x,t)]+...+\exp[p_{k}(x,t)]}\right],
\end{eqnarray}
where
$$p_{i}(x,t)=-a_{i}\frac{r_{3}(t_{0})}{r_{3}(t)}x+\frac{a_{i}^{2}}{2}\tau_{3}(t)+p_{i}^{0},\,\,\,\,\,\,\,\,\,\,\,\,i=1,2,...,k,$$
and $r_{3}(t)$ is given by  (\ref{x3}), $\tau_{3}(t)$ is given by (\ref{a3,tau3}).
 
\textbf{b.} Generalized triangular wave solution of BE (\ref{cal-kanai}) is of the form
(\ref{sim-solu}) with $r(t)=r_{3}(t)$ and $\tau(t)=\tau_{3}(t),$ $t>0.$

\textbf{c.} Generalized \emph{N}-wave  solution of  BE (\ref{cal-kanai}) for the over damping case can be written using formula (\ref{N-wave}).

\textbf{d.} The forced BE (\ref{cal-kanai}) with initial conditions $U_{m}(x,0)=-(m/x),\,\,\,\,\,m=1,2,3...,$
has rational type solutions of the form
\begin{equation*}\label{gu-33}
U_{m}(x,t)=\frac{\dot{r}_{3}(t)}{r_{3}(t)}x-e^{-\gamma t} (\frac{r_{3}(0)}{r_{3}(t)})\left(\frac{m H_{m-1}\left(\frac{r_{3}(0)}{r_{3}(t)}x,\frac{1}{2}\tau_{3}(t)\right)}
{H_{m}\left(\frac{r_{3}(0)}{r_{3}(t)}x,\frac{1}{2}\tau_{3}(t)\right)}\right),\,\,\,\,\,m=1,2,3,...,
\end{equation*}
where $r_{3}(t)$ and $\tau_{3}(t)$ are given by (\ref{x3}) and (\ref{a3,tau3}), respectively.

 From the general form of the solution (\ref{sol-U-3}), clearly if  the function $\,V(\eta(x,t),\tau(t)),\,$ is bounded for  $t\gg 1,$ then the long-time behavior of $U(x,t)$ is described by $\lim_{t\rightarrow \infty}U(x,t)=(-\gamma/2+\Omega')x,$ and the limiting function is a stationary solution of the BE (\ref{cal-kanai}).

\subsubsection{Variable coefficient case with $\omega_{0}^{2}<0.$} In the study of damped harmonic oscillator, usually $\gamma$ and $\omega_{0}$ are positive parameters  leading to the critical, under and the over damping cases, which we already discussed. However, we can take also $\,\omega_{0}^2=-\tilde{\omega}^2,$ $\tilde{\omega}>0,\,\gamma>0,$ and consider the Burgers equation
\begin{eqnarray}\label{cal-kanai-2}
 U_{t}+\gamma U+U U_{x}=\frac{1}{2}e^{-\gamma t}U_{xx}+\tilde{\omega}^{2}x\,\,,\,\,\,\,\,\,\,\,\,-\infty<x<\infty,\,\,\,\,t>0.
 \end{eqnarray}
In that case, we have the IVP $\,
\ddot{r}+\gamma \dot{r}-\tilde{\omega}^{2}r=0,\,\,r(0)=r_{0}\neq
0,\,\,\,\dot{r}(0)=0,\,$ which has solution
\begin{eqnarray*}
r(t)=r_{0}\frac{\tilde{\omega}}{\tilde{\Omega}}e^{-\frac{\gamma t}{2}}\cosh[\tilde{\Omega}t+\tilde{\beta}],
\end{eqnarray*}
where $\tilde{\Omega}=\sqrt{\tilde{\omega}^{2} +(\gamma^2/4)},$ and $\tilde{\beta}=\tanh^{-1}(\gamma/2\tilde{\Omega}).$
Then, BE (\ref{cal-kanai-2}) has solutions of the form,
\begin{eqnarray}\label{solu-w}
U(x,t)&=&(-(\gamma/2)+\tilde{\Omega}\tanh(\tilde{\Omega}t+\tilde{\beta}))x  \nonumber\\
& &+\frac{\tilde{\Omega}}{\tilde{\omega}}\left(\frac{e^{-\gamma t/2}}{\cosh(\tilde{\Omega}t+\tilde{\beta})}\right)
V\left(\frac{\tilde{\Omega}}{\tilde{\omega}}\left(\frac{e^{\gamma t/2}}{\cosh(\tilde{\Omega}t+\tilde{\beta})}\right),
\frac{\tilde{\Omega}}{\tilde{\omega}^2}\left(\tanh[\tilde{\Omega} t +\tilde{\beta}]-\frac{\gamma}{2\tilde{\Omega}}\right)\right).
\end{eqnarray}
In that case, when $\gamma\rightarrow 0,$ one has $\tilde{\Omega}\rightarrow \tilde{\omega},$ and $\tilde{\beta}\rightarrow 0.$ If $V(\eta,\tau)$ is as in the previous cases, then one can see that the solution  (\ref{solu-w}) of the variable parametric BE (\ref{cal-kanai-2}) approaches the solution (\ref{solu-1}) of the constant coefficient forced BE (\ref{st-bur-2}), which in turn approaches the standard BE, when $\tilde{\omega}\rightarrow 0$.

\section{Conclusion}

Nonlinear PDE's are important tool to study many physical and natural phenomena. The reach and complicated structure of their solutions corresponds to the reach and complicated character of the real world systems. In general, it is impossible to solve analytically a given nonlinear PDE. Only special class of integrable models admit exact solutions and these models play crucial role on revealing the nature of many physical phenomena, as well as provide convenient schemes to develop perturbation theory and test numerical methods.

In this article, we introduced exactly solvable nonhomogeneous Burgers equation with specific time variable coefficients, and analytic solution of the initial value problem was obtained in terms of a corresponding second order linear differential equation.  Burgers equations with constant damping, exponentially decaying diffusion coefficient and a forcing term linear in space variable were studied as particular cases. Generalized shock solitary waves, triangular waves, \emph{N}-waves and rational type solutions were found explicitly and graphically illustrated.
Special properties like  fusion of shocks in traveling wave solutions, and motion of poles of rational type solutions were observed.  In addition, we shortly discussed
the limiting case of the parametric equations, and the long-time behavior of their solutions.
 We remark that, our results can be applied also to study a wide class of variable parametric Burgers and KPZ equations related with the classical Sturm-Liouville problems for the orthogonal polynomials, \cite{S.O.E}.

Finally, we note that, there are different approaches to study the variable parametric Burgers  problems posed in this article. The one, which we used here, is transforming the nonhomogeneous Burgers equation with variable coefficients to a  standard Burgers equation, and then applying Cole-Hopf linearization.  Another approach is a direct linearization of the variable parametric Burgers equation in the form of a variable parametric heat equation, which in turn can be transformed to a standard heat equation or can be solved using the evolution operator method. These problems are under consideration.

\vspace{.2in}

\textbf{Acknowledgments:} This work is supported by the National Science Foundation of Turkey, T\"{U}BITAK, TBAG Project No: 110T679.

\end{document}